\documentclass[twocolumn]{aastex631}

\newcommand{\A}{\text{\normalfont\AA}}

\def \halpha{H$\alpha$}

\def \h2{{\rm H_{2}}}

\def \halpha{H$\alpha$}

\def \dn4000{D_{{\rm n}}(4000) }

\begin{document}

\title{COSMOS-Web: The Role of Galaxy Interactions and Disk Instabilities\\in Producing Starbursts at $z<4$}

\author[0000-0002-9382-9832]{Andreas L. Faisst}
\affiliation{Caltech/IPAC, 1200 E. California Blvd. Pasadena, CA 91125, USA}
\correspondingauthor{A. L. Faisst}
\email{afaisst@caltech.edu}


\author[0000-0002-8434-880X]{Lilan Yang}
\affiliation{Laboratory for Multiwavelength Astrophysics, School of Physics and Astronomy, Rochester Institute of Technology, 84 Lomb Memorial Drive, Rochester, NY 14623, USA}

\author[0000-0002-0245-6365]{M. Brinch}
\affiliation{Cosmic Dawn Center (DAWN), Denmark} 
\affil{DTU Space, Technical University of Denmark, Elektrovej 327, 2800 Kgs. Lyngby, Denmark}

\author[0000-0002-0930-6466]{C. M. Casey}
\affiliation{The University of Texas at Austin, 2515 Speedway Blvd Stop C1400, Austin, TX 78712, USA}
\affiliation{Cosmic Dawn Center (DAWN), Denmark} 

\author[0000-0003-3691-937X]{N. Chartab}
\affiliation{The Observatories of the Carnegie Institution for Science, 813 Santa Barbara St., Pasadena, CA 91101, USA}

\author[0000-0003-0348-2917]{M. Dessauges-Zavadsky}
\affiliation{D\'epartement d'Astronomie, Universit\'e de Gen\`eve, Chemin Pegasi 51, CH-1290 Versoix, Switzerland}

\author[0000-0003-4761-2197]{N. E. Drakos}
\affiliation{Department of Physics and Astronomy, University of Hawaii, Hilo, 200 W Kawili St, Hilo, HI 96720, USA}

\author[0000-0001-9885-4589]{S. Gillman}
\affiliation{Cosmic Dawn Center (DAWN), Denmark} 
\affiliation{DTU Space, Technical University of Denmark, Elektrovej 327, 2800 Kgs. Lyngby, Denmark}

\author[0000-0002-0236-919X]{G. Gozaliasl}
\affiliation{Department of Computer Science, Aalto University, P. O. Box 15400, Espoo, FI-00 076, Finland}
\affiliation{Department of Physics, University of Helsinki, P.O. Box 64, FI-00014 Helsinki, Finland}

\author[0000-0003-4073-3236]{C. C. Hayward}
\affiliation{Center for Computational Astrophysics, Flatiron Institute, 162 Fifth Avenue, New York, NY 10010, USA}

\author[0000-0002-7303-4397]{O. Ilbert}
\affiliation{Aix Marseille Univ, CNRS, CNES, LAM, Marseille, France  }

\author[0000-0002-9655-1063]{P. Jablonka}
\affil{Ecole Polytechnique F\'ed\'erale de Lausanne, CH-1015 Lausanne, Switzerland}

\author[0009-0000-3672-0198]{A. Kaminsky}
\affiliation{Department of Physics, University of Miami, Coral Gables, FL 33124, USA} 

\author[0000-0001-9187-3605]{J. S. Kartaltepe}
\affiliation{Laboratory for Multiwavelength Astrophysics, School of Physics and Astronomy, Rochester Institute of Technology, 84 Lomb Memorial Drive, Rochester, NY 14623, USA}

\author[0000-0002-6610-2048]{A. M. Koekemoer}
\affiliation{Space Telescope Science Institute, 3700 San Martin Dr., Baltimore, MD 21218, USA} 

\author[0000-0002-5588-9156]{V. Kokorev}
\affiliation{Kapteyn Astronomical Institute, University of Groningen, PO Box 800, 9700 AV Groningen, The Netherlands}

\author[0000-0003-3216-7190]{E. Lambrides}\altaffiliation{NPP Fellow}
\affiliation{NASA-Goddard Space Flight Center, Code 662, Greenbelt, MD, 20771, USA}

\author[0000-0001-9773-7479]{D. Liu}
\affiliation{Purple Mountain Observatory, Chinese Academy of Sciences, 10 Yuanhua Road, Nanjing 210023, China}

\author[0000-0001-7711-3677]{C. Maraston}
\affiliation{Institute of Cosmology and Gravitation, University of Portsmouth, Dennis Sciama Building, Burnaby Road, Portsmouth, PO13FX, United Kingdom}

\author[0000-0001-9189-7818]{C. L. Martin}
\affil{Department of Physics, University of California, Santa Barbara, Santa Barbara, CA 93109, USA}

\author[0000-0002-7093-7355]{A. Renzini}
\affiliation{INAF, Osservatorio Astronomico di Padova, Vicolo dell’Osservatorio 5, 35122, Padova, Italy}

\author[0000-0002-4271-0364]{B. E. Robertson}
\affiliation{Department of Astronomy and Astrophysics, University of California, Santa Cruz, 1156 High Street, Santa Cruz, CA 95064, USA}

\author[0000-0002-1233-9998]{D. B. Sanders}
\affiliation{Institute for Astronomy, University of Hawai’i at Manoa, 2680 Woodlawn Drive, Honolulu, HI 96822, USA}

\author[0000-0002-0364-1159]{Z. Sattari}
\affiliation{Department of Physics and Astronomy, University of California, Riverside, 900 University Ave, Riverside, CA 92521, USA}
\affiliation{The Observatories of the Carnegie Institution for Science, 813 Santa Barbara St., Pasadena, CA 91101, USA}

\author[0000-0002-0438-3323]{N. Scoville}
\affiliation{Astronomy Dept., California Institute of Technology, 1200 E. California Blvd, Pasadena, CA, USA}

\author[0000-0002-0745-9792]{C. M. Urry}
\affiliation{Physics Department and Yale Center for Astronomy \& Astrophysics, Yale University, PO Box 208120, CT 06520-8120, USA}

\author[0000-0002-1905-4194]{A. P. Vijayan}
\affiliation{Cosmic Dawn Center (DAWN), Denmark} 
\affiliation{DTU Space, Technical University of Denmark, Elektrovej 327, 2800 Kgs. Lyngby, Denmark}

\author[0000-0003-1614-196X]{J. R. Weaver}
\affil{Department of Astronomy, University of Massachusetts, Amherst, MA 01003, USA}

\author[0000-0003-3596-8794]{H. B. Akins}
\affiliation{The University of Texas at Austin, 2515 Speedway Blvd Stop C1400, Austin, TX 78712, USA}

\author[0000-0001-9610-7950]{N. Allen}
\affiliation{Cosmic Dawn Center (DAWN), Denmark} 
\affiliation{Niels Bohr Institute, University of Copenhagen, Jagtvej 128, DK-2200, Copenhagen, Denmark}

\author[0000-0002-0569-5222]{R. C. Arango-Toro}
\affiliation{Aix Marseille Univ, CNRS, CNES, LAM, Marseille, France  }

\author[0000-0003-3881-1397]{O. R. Cooper}\altaffiliation{NSF Graduate Research Fellow}
\affiliation{The University of Texas at Austin, 2515 Speedway Blvd Stop C1400, Austin, TX 78712, USA}

\author[0000-0002-3560-8599]{M. Franco}
\affiliation{The University of Texas at Austin, 2515 Speedway Blvd Stop C1400, Austin, TX 78712, USA}

\author[0000-0002-8008-9871]{F. Gentile}
\affiliation{University of Bologna - Department of Physics and Astronomy ``Augusto Righi'' (DIFA), Via Gobetti 93/2, I-40129 Bologna, Italy}
\affiliation{INAF, Osservatorio di Astrofisica e Scienza dello Spazio, Via Gobetti 93/3, I-40129, Bologna, Italy}

\author[0000-0003-0129-2079]{S. Harish}
\affiliation{Laboratory for Multiwavelength Astrophysics, School of Physics and Astronomy, Rochester Institute of Technology, 84 Lomb Memorial Drive, Rochester, NY 14623, USA}

\author[0000-0002-3301-3321]{M. Hirschmann}
\affiliation{Institute of Physics, GalSpec, Ecole Polytechnique F\'ed\'erale de Lausanne, Observatoire de Sauverny, Chemin Pegasi 51, 1290 Versoix, Switzerland}
\affiliation{INAF, Astronomical Observatory of Trieste, Via Tiepolo 11, 34131 Trieste, Italy}

\author[0000-0002-0101-336X]{A. A. Khostovan}
\affiliation{Laboratory for Multiwavelength Astrophysics, School of Physics and Astronomy, Rochester Institute of Technology, 84 Lomb Memorial Drive, Rochester, NY 14623, USA}
\affiliation{Astrophysics Division, NASA Goddard Space Flight Center, Greenbelt, MD, 20771, USA}

\author[0009-0008-5926-818X]{C. Laigle}
\affiliation{Institut d’Astrophysique de Paris, UMR 7095, CNRS, and Sorbonne Universit\'e, 98bis boulevard Arago, 75014 Paris, France}

\author[0000-0003-2366-8858]{R. L. Larson}
\altaffiliation{NSF Graduate Fellow}
\affiliation{The University of Texas at Austin, 2515 Speedway Blvd Stop C1400, Austin, TX 78712, USA}

\author[0000-0002-2419-3068]{M. Lee}
\affiliation{Cosmic Dawn Center (DAWN), Denmark} 
\affiliation{DTU Space, Technical University of Denmark, Elektrovej 327, 2800 Kgs. Lyngby, Denmark}

\author[0000-0002-9252-114X]{Z. Liu}
\affiliation{Kavli Institute for the Physics and Mathematics of the Universe (WPI), The University of Tokyo, Kashiwa, Chiba 277-8583, Japan}
\affiliation{Center for Data-Driven Discovery, Kavli IPMU (WPI), UTIAS, The University of Tokyo, Kashiwa, Chiba 277-8583, Japan}
\affiliation{Department of Astronomy, School of Science, The University of Tokyo, 7-3-1 Hongo, Bunkyo, Tokyo 113-0033, Japan}

\author[0000-0002-7530-8857]{A. S. Long}\altaffiliation{NASA Hubble Fellow}
\affiliation{The University of Texas at Austin, 2515 Speedway Blvd Stop C1400, Austin, TX 78712, USA}

\author[0000-0002-4872-2294]{G. Magdis}
\affiliation{Cosmic Dawn Center (DAWN), Denmark} 
\affiliation{Niels Bohr Institute, University of Copenhagen, Jagtvej 128, DK-2200, Copenhagen, Denmark}
\affiliation{DTU Space, Technical University of Denmark, Elektrovej 327, 2800 Kgs. Lyngby, Denmark}

\author[0000-0002-6085-3780]{R. Massey}
\affil{Department of Physics, Centre for Extragalactic Astronomy, Durham University, South Road, Durham DH1 3LE, UK}

\author[0000-0002-9489-7765]{H. J. McCracken}
\affiliation{Institut d’Astrophysique de Paris, UMR 7095, CNRS, and Sorbonne Université, 98 bis boulevard Arago, F-75014 Paris, France}

\author[0000-0002-6149-8178]{J. McKinney}
\affiliation{The University of Texas at Austin, 2515 Speedway Blvd Stop C1400, Austin, TX 78712, USA}

\author[0000-0003-2397-0360]{L. Paquereau} 
\affiliation{Institut d’Astrophysique de Paris, UMR 7095, CNRS, and Sorbonne Université, 98 bis boulevard Arago, F-75014 Paris, France}

\author[0000-0002-4485-8549]{J. Rhodes}
\affiliation{Jet Propulsion Laboratory, California Institute of Technology, 4800 Oak Grove Drive, Pasadena, CA 91001, USA}

\author[0000-0003-0427-8387]{R. M. Rich}
\affiliation{Department of Physics and Astronomy, UCLA, PAB 430 Portola Plaza, Box 951547, Los Angeles, CA 90095-1547}

\author[0000-0002-7087-0701]{M. Shuntov}
\affiliation{Cosmic Dawn Center (DAWN), Denmark} 
\affiliation{Niels Bohr Institute, University of Copenhagen, Jagtvej 128, DK-2200, Copenhagen, Denmark}

\author[0000-0002-0000-6977]{J. D. Silverman}
\affiliation{Kavli Institute for the Physics and Mathematics of the Universe (WPI), The University of Tokyo, Kashiwa, Chiba 277-8583, Japan}
\affiliation{Department of Astronomy, School of Science, The University of Tokyo, 7-3-1 Hongo, Bunkyo, Tokyo 113-0033, Japan}

\author[0000-0003-4352-2063]{M. Talia}
\affiliation{University of Bologna - Department of Physics and Astronomy ``Augusto Righi'' (DIFA), Via Gobetti 93/2, I-40129 Bologna, Italy}
\affiliation{INAF, Osservatorio di Astrofisica e Scienza dello Spazio, Via Gobetti 93/3, I-40129, Bologna, Italy}

\author[0000-0003-3631-7176]{S. Toft}
\affiliation{Cosmic Dawn Center (DAWN), Denmark} 
\affiliation{Niels Bohr Institute, University of Copenhagen, Jagtvej 128, DK-2200, Copenhagen, Denmark}

\author[0000-0002-7051-1100]{J. A. Zavala}
\affiliation{National Astronomical Observatory of Japan, 2-21-1 Osawa, Mitaka, Tokyo 181-8588, Japan}



\begin{abstract}

We study of the role of galaxy-galaxy interactions and disk instabilities in producing starburst activity in galaxies out to $z=4$. For this, we use a sample of 387 galaxies with robust total star formation rate measurements from Herschel, gas masses from ALMA, stellar masses and redshifts from multi-band photometry, and JWST/NIRCam rest-frame optical imaging.
Using mass-controlled samples, we find an increased fraction of interacting galaxies in the starburst regime at all redshifts out to $z=4$. This increase correlates with star formation efficiency (SFE), but not with gas fraction. However, the correlation is weak (and only significant out to $z=2$), which could be explained by the short duration of SFE increase during interaction.
In addition, we find that isolated disk galaxies make up a significant fraction of the starburst population. The fraction of such galaxies with star-forming clumps (``clumpy disks'') is significantly increased compared to the main-sequence disk population. Furthermore, this fraction directly correlates with SFE. This is direct observational evidence for a long-term increase of SFE maintained due to disk instabilities, contributing to the majority of starburst galaxies in our sample and hence to substantial mass growth in these systems. This result could also be of importance for explaining the growth of the most massive galaxies at $z>6$.
\end{abstract}



\section{Introduction} \label{sec:intro}

For most of their lives, star-forming galaxies follow the main-sequence \citep[e.g.,][]{noeske07,daddi07,elbaz07}, a tight relation between stellar mass and star formation rate (SFR) set by an equilibrium state between gas consumption, outflow, and inflow \citep[][]{dave12,lilly13,feldmann15}. The scatter of this relation ($\sim0.3\,{\rm dex}$) is set by oscillations around that equilibrium, driven by constant adjustment of the galaxies' SFR to the available cold gas and replenishing gas inflows \citep[e.g.,][]{tacchella16,dekel13,bouche10}. Stronger oscillations are observed mostly in low-mass galaxies at low and local redshifts, while at higher redshifts galaxies across all masses seem to show such a behavior \citep[e.g.,][]{weisz12,emami19,faisst19}.
In addition to these oscillations, galaxies experience significant bursts of star formation, elevating their SFRs by factors of $10$ and more above the main sequence \citep{sanders96}. The fraction of starburst galaxies to the total population increases with redshift from $\sim1\%$ at $z<0.4$ \citep{bergvall16} to $\sim 2-5\%$ at $z\sim 0.5-1.0$ \citep{bisigello18,rodighiero11} and higher fractions at higher redshifts depending on stellar mass \citep[][]{sargent12,caputi17,bisigello18}.
Interactions between gas-rich galaxies lead to the inflow and compression of gas and are, therefore, usually the prerequisites for starbursts \citep[e.g.,][]{sanders96,cox08,genzel10,kartaltepe12,hung13,hopkins18,cibinel19,moreno21,shah22}. However, some simulations and observations suggest that starbursts and galaxy major mergers do not have to be necessarily causally connected \citep[e.g.,][]{rodighiero11,kaviraj13,sparre16,fensch17,wilkinson22,li23,zliu24}.
In addition to starbursts, disk instabilities in gas-rich galaxies \citep{toomre64,bournaud07}, which may be tightly related to increased gas inflows or minor mergers \citep{scoville23}, could increase the rate of star formation without external interactions, hence contributing to the starburst population \citep[e.g.,][]{wang94,immeli04,tadaki18,dekel09} and the build-up of bulge-dominated galaxies later on \citep[e.g.,][]{bouwens99}.
The above shows that the origin of starbursts is, therefore, complex, and the definitive link between galaxy interactions, disk structure, increased star formation efficiency (SFE), and gas fraction out to high redshift has yet to be studied with robust measurements and comprehensive samples.

The works by \citet{liu19} and \citet{scoville23} are likely the most comprehensive sub-mm dust-based studies (in terms of sample size and redshift coverage) on the relation between gas and star formation properties of main-sequence and starburst galaxies to date. They combine robust SFRs from Herschel, gas fraction and SFE measurements from ALMA, and other parameters (such as stellar masses) from comprehensive multi-band photometry for $>700$ galaxies out to $z=6$. Specifically, \citet{scoville23} found that starbursts at a given redshift and stellar mass are primarily triggered due to enhanced SFE rather than a high gas fraction. Similar results have been found in independent studies by \citet{silverman15} and \citet{tacconi18} using large-sample CO data \citep[for a review see][]{tacconi20}. On the other hand, an increase in the gas fraction is likely responsible for the maintained high specific SFRs at high redshift \citep[e.g.,][]{tacconi10,genzel15,tacconi18,freundlich19,liu19,gowardhan19,dessauges20}.

In this work, we explore the population of starburst galaxies out to $z=4$ in terms of their interactions, disk structure, SFE, and gas fractions. Specifically, we aim to study the role of galaxy-galaxy interactions and disk instabilities in producing starburst activity. To this end, we combine the \citet{scoville23} sample with high-resolution rest-frame optical imaging from the COSMOS-Web JWST/NIRCam imaging \citep{casey23}.

After summarizing the data (Section~\ref{sec:data}), we detail the morphological classification in Section~\ref{sec:structure}. In Section~\ref{sec:results}, we present the results and we conclude in Section~\ref{sec:discussion}.
Throughout this work, we assume a $\Lambda$CDM cosmology with $H_0 = 70\,{\rm km\,s^{-1}\,Mpc^{-1}}$, $\Omega_\Lambda = 0.7$, and $\Omega_{\rm m} = 0.3$ and magnitudes are given in the AB system \citep{oke74}. We use a \citet{chabrier03} initial mass function (IMF) for stellar masses and SFRs.

\section{Sample and Basic Measurements} \label{sec:data}

For this work, we use a sample of 387 galaxies out to $z\sim4$. This sample is a sub-sample of the $704$ galaxies from \citet{scoville23} (see more details in earlier works; \citealt{scoville14,scoville16,scoville17}) requiring existing JWST imaging. The selected galaxy sample is best for carrying out this study as it includes
\begin{enumerate}
    \item robust measurements of total SFR from UV$+$far-IR Herschel and ALMA measurements; \vspace{-1mm}
    
    \item measured molecular gas masses from far-IR dust continuum observations with ALMA; \vspace{-1mm}
    
    \item measured stellar masses and accurate photometric redshifts from multi-band photometry; and \vspace{-1mm}
    
    \item deep sub-kpc resolution JWST rest-frame optical and near-IR imaging data.
\end{enumerate}

The galaxies reside in the COSMOS field \citep{scoville07}, which provides a wealth of ancillary data including X-ray and radio measurements, {\it Hubble} rest-frame UV and optical imaging \citep[ACS/F814W;][]{koekemoer07}, as well as spectroscopic redshifts for more than half of the sample. In the following, we summarize the different data products, basic measurements, and sample properties.

\subsection{Photometry and Imaging}

\paragraph{UV and Optical Photometry~}~
All photometric redshifts, UV-based SFRs, and stellar masses are based on the multi-band COSMOS2020 catalog photometry \citep{weaver22}. Obvious AGNs are removed from the sample by a combination of SED fitting (based on a comparison of the goodness of fit between star forming and AGN templates, see \citealt{weaver22}) and selection in X-ray and radio emission.
The {\it dust-unobscured} star formation is measured from the $1500\,{\rm \AA}$ emission constrained by the best-fit SED. Stellar masses are derived using \texttt{LePhare} \citep[][]{arnouts99,ilbert06}, assuming a variety of templates, ages, metallicities, and dust attenuations.
Uncertainties in SFRs and stellar masses are generally less than $0.3\,{\rm dex}$ \citep{weaver22}.
Spectroscopic redshifts are available for 58\% of the galaxies in our sample from various surveys (e.g., \citealt{lefevre15,hasinger18}; Khostovan et al. in prep.) and suggest a photometric redshift accuracy better than $\sigma_{z} = 0.14$.
Note that using the spectroscopic redshifts for deriving stellar masses and other properties changes their values within their uncertainties, thus to would not change the conclusions of this work.

\begin{figure*}[t!]
\centering
\includegraphics[angle=0,width=2.1\columnwidth]{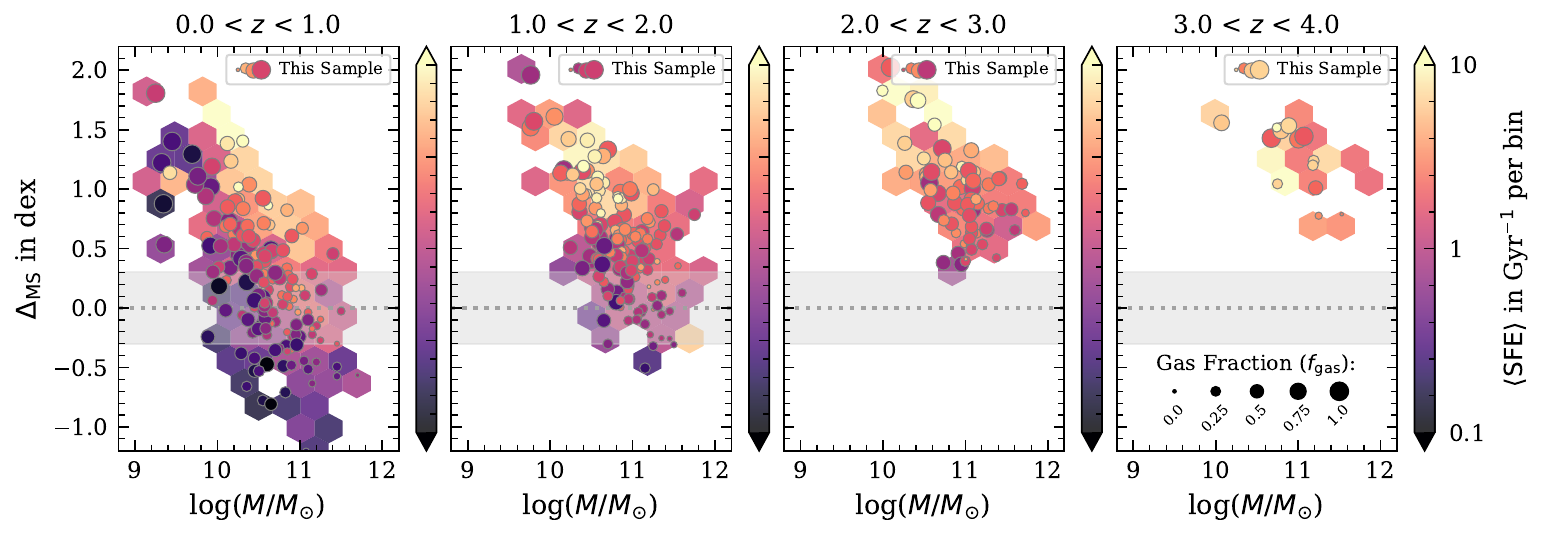}\\
\caption{Dependency of star formation efficiency (SFE; color-coded from $0.1\,{\rm Gyr^{-1}}$ [dark] to $10\,{\rm Gyr^{-1}}$ [light]) and molecular gas fraction ($f_{\rm gas}$; size of circles from $0$ [small] to $1$ [large]) on stellar mass and offset from the star-forming main sequence ($\Delta_{\rm MS}$) in our sample. Each panel shows a different redshift bin. The underlying hex bins show the SFE for the full sample of $704$ galaxies from \citet{scoville23}. Note the stronger correlation between $\Delta_{\rm MS}$ and SFE compared to $f_{\rm gas}$. Also visible is a selection effect due to the requirement of {\it Herschel} detections, which leads to the lack of galaxies on the main sequence at lower stellar masses (see text for more detailed discussion).
\label{fig:sampleprop}}
\end{figure*}

\paragraph{JWST Imaging (for morphology)~}~
JWST provides rest-frame optical/NIR imaging at $30\,{\rm mas}$ {\it pixel scale} from two cycle 1 GO programs: COSMOS-Web \citep[PID 1727; PIs: Kartaltepe \& Casey; ][]{casey23} covering $0.54\,{\rm deg^2}$ of the COSMOS field and PRIMER-COSMOS (PID 1837; PI: Dunlop) covering the CANDELS-COSMOS field. In this work we make use of the full COSMOS-Web dataset including the NIRCam F115W, F150W, F277W, and F444W filters over the full area. PRIMER-COSMOS uses F090W, F150W, F200W, F277W, F356W, F410M, and F444W. The PSF FWHM varies between $40\,{\rm mas}$ and $145\,{\rm mas}$ for the different filters. The physical scale varies from $6.2\,{\rm kpc/\arcsec}$ at $z=0.5$ to $7.4\,{\rm kpc/\arcsec}$ at $z=3.5$. MIRI imaging is not used here due to the small area coverage and shallow depth. See \citet{franco23} for the details of the JWST data reduction. These rest-frame optical/NIR JWST observations are crucial to trace the bulk of stellar mass, as rest-frame UV imaging would be biased to unobscured star-forming regions in the galaxies and therefore could mislead the identification of merging system \citep[see][]{cibinel19}. (However, we note that rest-frame UV imaging is used to identify galaxies with star-forming clumps.)
We note that the JWST photometry is not used for SED fitting as it does not add much in terms of wavelength coverage and depth to the available ground-based data for these relatively bright galaxies. However, we checked internally that the photometry between JWST and ground-based observations is consistent and we therefore do not expect any biases in the photometry.

\paragraph{Herschel Photometry and Measurements~}~
Infrared photometry is derived from various data available for the COSMOS field: at $24\,{\rm \mu m}$ by Spitzer \citep{sanders07}; at $100\,{\rm \mu m}$ and $160\,{\rm \mu m}$ by Herschel-PACS \citep[][]{poglitsch10} as part of the {\it PACS Evolutionary Probe} program \citep[PEP; ][]{lutz11}; at $250\,{\rm \mu m}$, $350\,{\rm \mu m}$, and $500\,{\rm \mu m}$ by Herschel-SPIRE \citep{griffin10} as part of the {\it Herschel Multi-tiered Extragalactic Survey} \citep[HerMES; ][]{oliver12}.
For the flux extraction a linear inversion technique of cross-identification \citep[``XID'',][]{roseboom10,roseboom12} is used based on positional priors from the Spitzer $24\,{\rm \mu m}$ catalog and VLA $1.4\,{\rm GHz}$ data \citep{lefloc09,schinnerer10}. All sources are detected at $>3\sigma$ in at least two of the five Herschel bands. For detailed information on de-blending algorithms we refer to appendix C2 in \citet{scoville23} as well as \citet{lee13}.
The infrared SFRs are measured from the total infrared luminosity via ${\rm SFR_{IR}\,[M_\odot\,yr^{-1}]} = 8.6 \times 10^{-11}\,{\rm L_{IR}\,[L_\odot]}$. This assumes that all stellar light is dust obscured in the first $100\,{\rm Myrs}$ (and none thereafter). The derived infrared SFRs would increase by $50\%$ for dust-enshrouded time scales of $10\,{\rm Myrs}$ \citep{scoville13}. The infrared luminosity is measured by integrating over a modified black body \citep{casey12} fit to the infrared photometry (see \citealt{scoville23} for more details).
The infrared SFRs are combined with the UV SFRs to obtain the total star formation rates.

\paragraph{ALMA Continuum Measurements~}~
The molecular ISM gas masses are derived from the Rayleigh-Jeans (RJ) dust continuum (rest-frame $850\,{\rm \mu m}$) as described in \citet{scoville14}. The $L_{\rm \nu_{850\,{\mu m}}} / M_{\rm gas}$ ratio was calibrated over a range of galaxy types (main-sequence, starburst, luminous infrared galaxies) out to $z=3$.
The RJ continuum is derived from archival ALMA bands $6$ and $7$ observations (at $>2\sigma$ significance) as of June 2021. Only data with {\it uv} coverage resolving a source extend of $\sim 1\arcsec$ is used for robust flux measurements. The fluxes and their uncertainties are derived from least-squares fitting. More information on the flux measurements is provided in \citet{scoville23}.

\begin{figure*}[t]
\centering
\includegraphics[angle=0,width=0.98\textwidth]{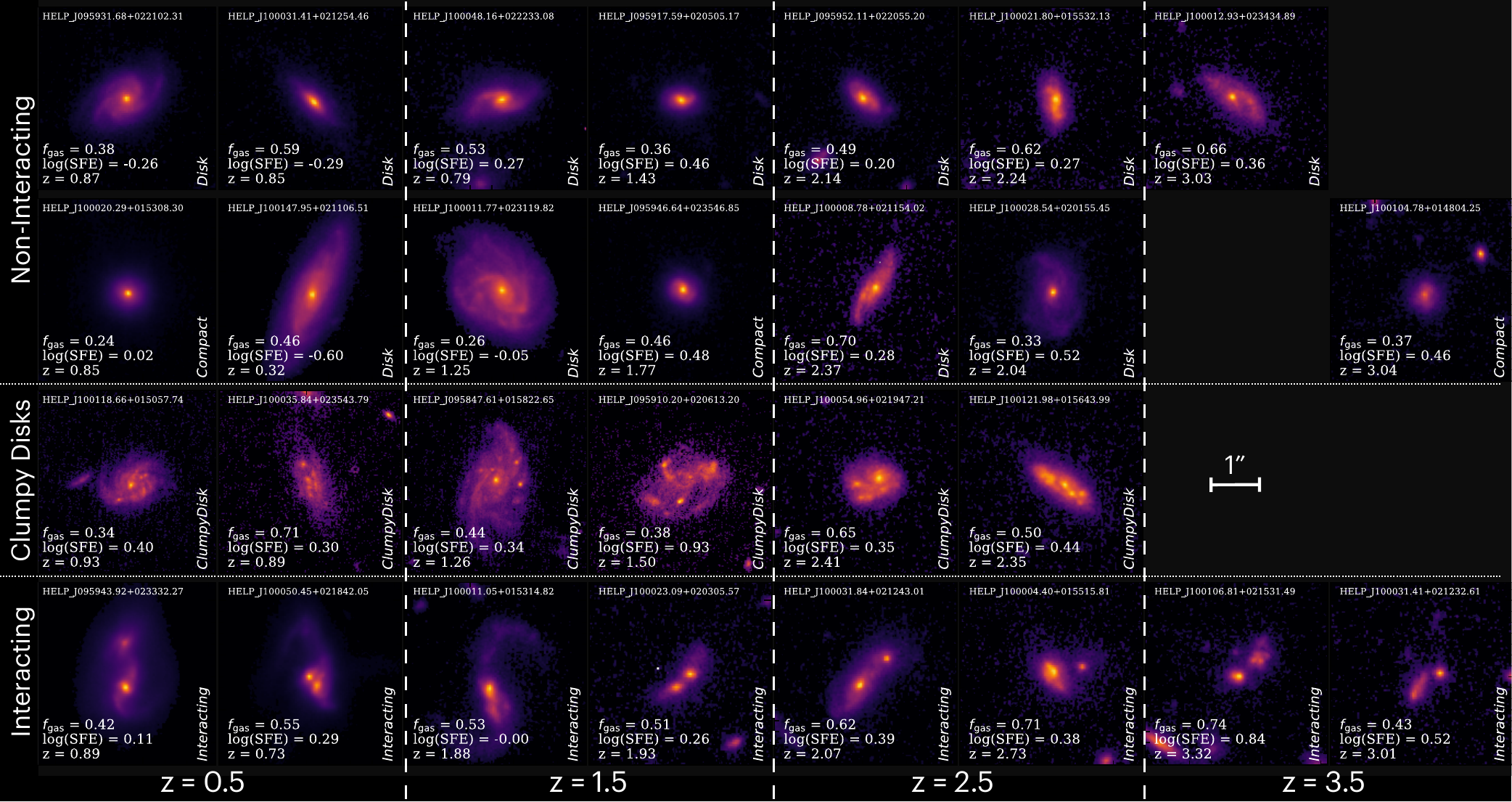}
\caption{Examples of galaxies imaged in JWST NIRCam/F277W at different redshifts organized in three different morphological groups (here {\it Disk} galaxies are combined in the {\it Non-Interacting} group). Sub-categories (such as {\it Disk} vs. {\it Clumpy Disk} vs. {\it Compact}) are labeled in the cutouts as well as $f_{\rm gas}$, SFE, and redshifts. Note that we do not classify clumpy disks at $z>3$ due to lack of resolution and significant confusion with interactions.
\label{fig:morph}}
\end{figure*}

\subsection{Definitions and Sample Properties}\label{sec:definitions}

We define the molecular gas fraction as $f_{\rm gas} = M_{\rm mol,gas} / (M_{\rm *} + M_{\rm mol,gas})$ and the star-formation efficiency based on the total UV$+$IR SFR as ${\rm SFE} = M_{\rm mol,gas} / {\rm SFR}$ in units of $\rm Gyr^{-1}$. The depletion time ($t_{\rm depl}$) is the inverse of the SFE. To calculate $\rm \Delta_{\rm MS}$ (the logarithmic offset from the star-forming main-sequence at a given stellar mass), we assume the main-sequence parameterization from \citet{lee15}, which is based on Herschel measurements of individual galaxies and stacked samples on the COSMOS field. We emphasize that Herschel observations are crucial for deriving robust $\rm \Delta_{\rm MS}$ of starburst galaxies, often dominated by dust-obscured star formation.
The use of other parameterizations \citep[e.g.,][]{speagle14,schreiber15} lead to similar $\Delta_{\rm MS}$ and have a minimal impact on the final results of this work. 

Figure~\ref{fig:sampleprop} shows the dependence of SFE and $f_{\rm gas}$ on stellar mass and $\Delta_{\rm MS}$ for our selected 387 galaxies in different redshift bins. The underlying hex bins show the full sample from \citet{scoville23}. This shows that our selected sub-sample traces the full sample properties closely. In addition, note that $\Delta_{\rm MS}$ correlates well with SFE and less with $f_{\rm gas}$, as it has been found in various works including that of \citet{silverman15} and \citet{scoville14}.

There are mainly two selection biases to be noted. These are due to the requirement of {\it Herschel} and ALMA detections for the computation of accurate far-IR SFRs and gas masses. First, the high-redshift sample is biased towards high $\Delta_{\rm MS}$. Second, low stellar masses are biased towards high $\Delta_{\rm MS}$. While the first bias is not of a large concern (our analysis will be done in separate redshift bins), the second bias may introduce unphysical relations with stellar mass. We therefore mitigate that bias by introducing mass-controlled samples for the following analysis by adopting two mass bins for each redshift range.
We found that two stellar mass bins are sufficient to mitigate the bias and more mass bins would reduce the sample and decrease the statistical robustness. However, due to the small sample at $z>3$, we only adopt a single mass bin for the highest redshifts.
We define the mass cuts in each of the redshift bins based on Figure~\ref{fig:sampleprop} to remove any significant dependence between stellar mass and $\Delta_{\rm MS}$ within the mass-binned sub-samples. The adopted mass bins (in $\log(M/{\rm M_\odot})$) are [10.1,10.8] and [10.8,11.7] for $z\sim 0.5$, [10.4,11.1] and [11.1,11.8] for $z\sim1.5$, [10.7,11.1] and [11.1,11.7] for $z\sim2.5$, and [10.7,11.2] for $z\sim 3.5$ (all redshift bins with $\Delta z = 1$). Changing these mass bins in a reasonable range ($\pm 0.2\,{\rm dex}$) does not affect the final conclusions of this work.

\section{Structural Analysis} \label{sec:structure}

\subsection{Visual classification}\label{sec:visclass}
We perform a visual classification of the 387 galaxies using all of the COSMOS-Web JWST NIRCam filters to minimize the effects of color-dependent morphology corrections.
In the following, we define four morphological groups:

\begin{itemize}
    \item[] \textbf{\it \textbf{Non-Interacting.}} Isolated galaxies of various kinds, such as disk or compact galaxies, that are not in an interacting state.

    \item[] \textbf{\it \textbf{Disk Galaxies.}} These are extended galaxies (to be distinguished from compact galaxies), preferentially with a semi-minor to semi-major axis ratio of less than $0.8$ and a disk structure. The disk can either be face-on or edge-on.
    
    \item[] \textbf{\it \textbf{Clumpy Disks.}} This is a subset of the {\it Disk Galaxies} category, showing more than one star-forming clump. The clumps are pronounced in the bluer bands but visible throughout redder wavelengths. Due to limitations in resolution and confusion with interacting systems, we have not classified clumpy disks at redshifts $z>3$ in this work.
    
    \item[] \textbf{\it \textbf{Interacting.}} Interacting or irregular galaxy systems are identified by multiple nuclei, irregular structures, or tidal features. This category should include mainly galaxy systems in pre-merger and post-merger phases.
\end{itemize}

It is important to note that we must make use of all available NIRCam filters for a reliable morphological classification.
One example to showcase the importance of multiple photometric bands for a morphological classification is discussed in \citep{zliu24}. As shown in their figure 1, this starburst galaxy could be classified as clear merger in JWST filters F115W and F150W. However, at longer rest-frame wavelengths (specifically F227W), this source shows a clear disk structure -- a finding that is supported by kinematic measurement of the CO gas.

For consistency across redshifts, we therefore visually classified interacting galaxies primarily at rest-frame optical wavelengths ($0.8-1\,{\rm \mu m}$) and redder bands where available. This choice is motivated by the fact that those wavelengths trace the stellar mass of galaxies, hence are more ideal to discriminate between interacting systems and galaxies with a significant disk disk component as shown in the case by \citep{zliu24}. We use F150W for $z<1.0$, F277W for $1 < z < 3$, and F444W for $z>3$, but also consider redder bands where available.
In addition, RGB color images (in addition to redshifts where available) are used to identify companions (mergers) and reject background/foreground galaxies.
We note that the PSF varies significantly between those bands \citep[from FWHM $0.06\arcsec$ to $0.16\arcsec$ for F115W to F444W;][]{zhuang24}, which may be a concern for high redshifts. We tested our classification by carefully comparing PSF-matched images to rule out significant biases in our visual classification due to such PSF differences.
Although we made use of all available bands in the morphological classification, we additionally tested the dependence of our morphological classification on single bands. We find, as expected, that using filters at bluer wavelengths breaks up the galaxies in clumps (see also below) and biases the morphological classification against disk galaxies. On the other hand, the use of filters redder than F227W for $z > 1$ galaxies does not change the classification. In addition, $\sim 5\%$ of galaxies are covered by the PRIMER-COSMOS observations, which include four more filters in addition to COSMOS-Web. The morphological classification was also tested against these filters (specifically F356W providing and intermediate band between F277W and F444W at a better PSF resolution than F444W) and we do not find biases using redder bands for the classification of interacting systems.

On the other hand, we use bluer bands (F115W and F150W) for the identification of clumpy disk galaxies. This is motivated by the fact that such clumps are generally star forming and rest-UV bright (and at the same time the PSF of bluer bands is smaller). However, there may be a bias in clump identification in high-redshift galaxies caused by their smaller physical sizes and the large PSF at same rest-frame filters. Furthermore, a differentiation between clumpy disks and interacting galaxies without kinematic information from spectroscopy is nearly impossible in these cases \citep{jones21}. We therefore do not attempt to identify clumpy disk galaxies at $z > 3$.
In addition to the above method, we use residual images for identifying clumps. These are generated in two ways, {\it (i)} by subtracting a smooth profile fit (Sersic) performed using \texttt{statmorph} \citep{rodriguezgomez19} from the galaxy image, and {\it (ii)} by directly subtracting a long-wavelength from a short-wavelength image (e.g., F150W - F444W), which highlights blue rest-UV structure and removes the smooth ISM. We classify a galaxy as {\it clumpy disk} if there are more than three bright (${\rm S/N > 3}$) clumps detected. We use both method to support the visual classification of clumpy disks from the individual or RGB images.

Finally, we compare the above visual classification with more statistical classification such as the $G-{ M_{\rm 20}}$ (Gini vs. $M_{\rm 20}$) or $C-A$ (concentration vs. asymmetry) diagnostics. We find that these methods are most efficient in selecting interacting galaxies, however, they have a significantly lower purity and completeness for selecting clumpy disks (see Appendix~\ref{sec:statmorph}).

In the following, we denote the fraction of interacting galaxies in our sample with $f_{\rm int}$. Note that, as mentioned above, without spectroscopic confirmation of the detailed kinematics, we are not able to identify close-pair mergers. The {\it Interacting} group therefore mostly focuses on the pre- and post-merger phases.
The fraction of clumpy disk galaxies compared to the total disk galaxy population in a given selection bin is denoted as $f_{\rm clumpy}$.
We derive the uncertainties on these fractions using the formalism of (Bayesian) binomial confidence intervals suggested by \citet{cameron11} and shown in code-form in their appendix A.

Figure~\ref{fig:morph} shows examples of our classification scheme in four redshift bins. The redshifts and galaxy types are indicated on each cutout, as well as $f_{\rm gas}$ and SFE. Note the clear distinction of disk galaxies with and without clumps.

\begin{figure*}[t]
\centering
\includegraphics[angle=0,width=0.95\textwidth]{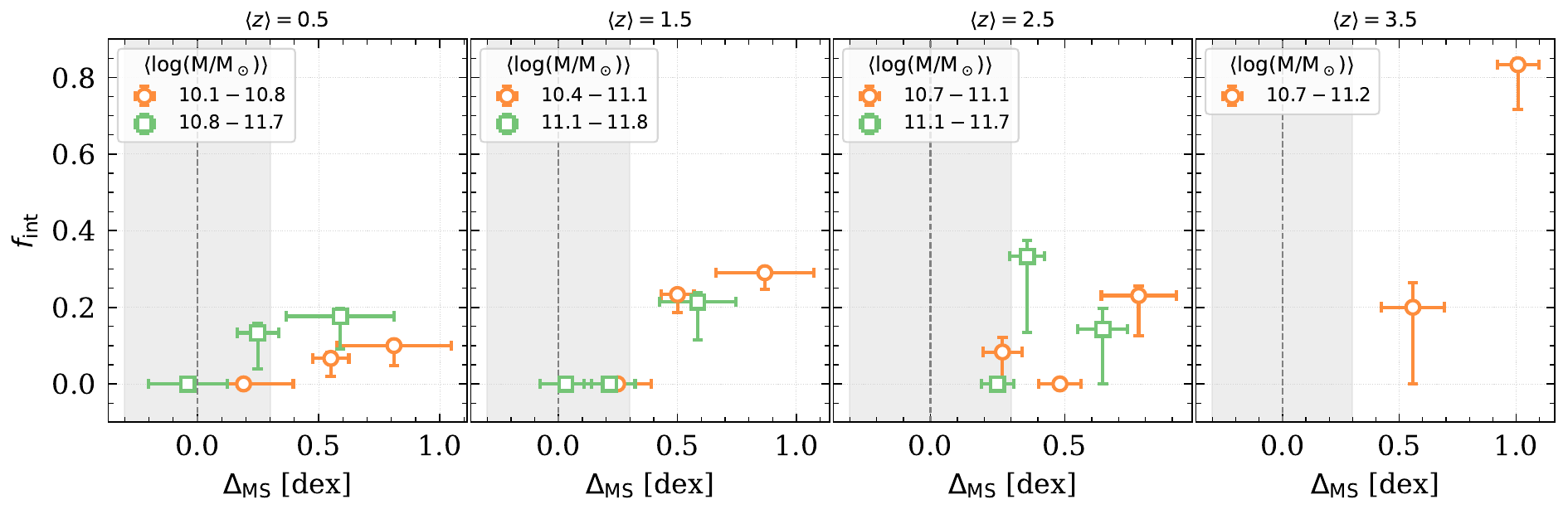}\\
\includegraphics[angle=0,width=0.95\textwidth]{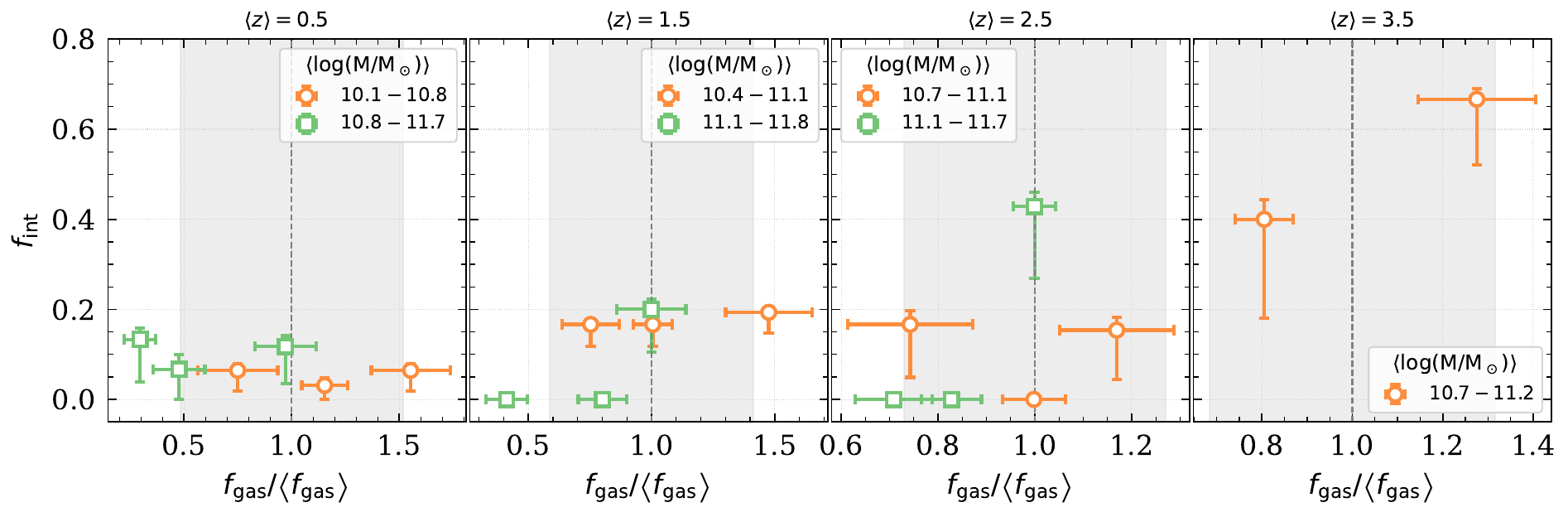}\\
\includegraphics[angle=0,width=0.95\textwidth]{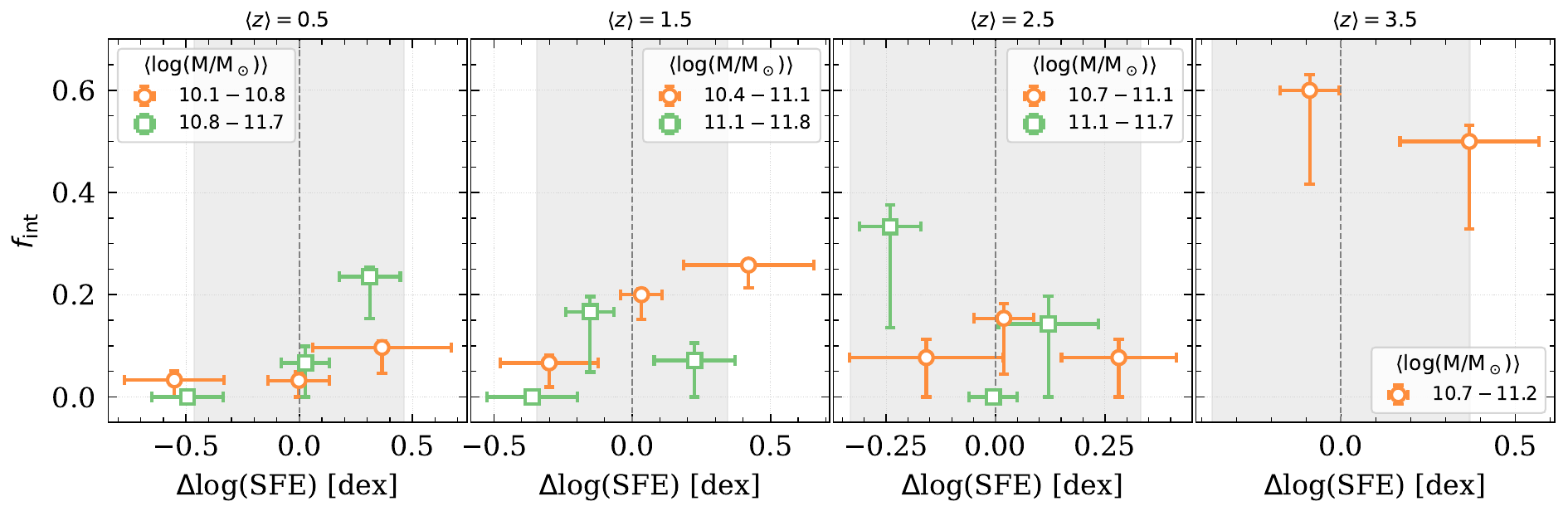}
\vspace{-3mm}
\caption{Relation between the fraction of interacting systems ($f_{\rm int}$) and the offset from the main sequence ($\Delta_{\rm MS}$; top), gas fraction ($f_{\rm gas}$, middle), and star formation efficiency (SFE; bottom) for different redshift ranges ($\Delta z = 1$) and two stellar mass bins. The latter two quantities are normalized to the mean of the population (width indicated by the gray region) at a given redshift and stellar mass.
\label{fig:fint}}
\end{figure*}

\section{Results} \label{sec:results}

\subsection{The Impact of Galaxy-Galaxy Interactions on Starburst Activity}

We first compare the fraction of interacting systems to the offset from the main sequence, the gas fraction, and the star formation efficiency.

The top panel in Figure~\ref{fig:fint} shows the fraction of interacting galaxies as a function of $\Delta_{\rm MS}$ in two different mass bins per redshift.
The fraction clearly increases off the main sequence (indicated by the gray area) towards higher $\Delta_{\rm MS}$ and reaches $20\%-40\%$ at $z<3$ and $80\%$ at $z=3.5$.
However, we note that, due to the lack of near-IR observations at these redshifts, the merger classification may not be as robust as at lower redshifts (for example due to dust obscuration for starburst galaxies). This can lead to a classification bias (e.g., due to clumpy disks, see discussion in Section~\ref{sec:visclass}) that may lead to an overprediction of the fraction of merging galaxies at $z=3.5$.

Note that here we focus change in fractions across different properties and do not attempt to compare their absolute values to the literature.
This is because relative changes are more reliable than the absolute values of $f_{\rm int}$, which depend on sample selection and on subjective visual selection thresholds and therefore are difficult to compare. The two mass bins show very similar behaviors. Keeping this in mind, an increase of the fraction of interacting systems towards higher redshifts is found, which is in agreement with a global increase in the merger fraction studied in other works \citep[e.g.,][]{tasca14,romano21}.
Overall, this is consistent with the theory of galaxy-galaxy interactions playing some role in inducing starburst activity. Here we show that this may be the case out to $z\sim4$. We also note that this is in line with and directly related to the increased fraction of interacting galaxies found at higher infrared luminosities \citep[cf. study at $z<1.5$ by][]{hung13}.

It is suggested that gas fraction is weakly correlated to starburst activity \citep[e.g.,][]{scoville23} and we therefore would not expect a significant correlation between $f_{\rm int}$ and $f_{\rm gas}$. The middle panels of Figure~\ref{fig:fint} show that this is indeed the case by displaying $f_{\rm int}$ as a function gas fraction of galaxies relative to the population mean ({\it i.e.} for a given stellar mass and redshift). No significant correlation between those quantities is seen in either stellar mass bins.

Finally, the lower panels in Figure~\ref{fig:fint} show $f_{\rm int}$ as a function of SFE (relative to the mean of the population at a given redshift and stellar mass). Up to $z=2$, we find a clear trend of galaxies with a higher-than-average SFE residing more frequently in interacting systems, specifically in the lower mass bin.
At higher redshifts, this trend is less significant. As discussed in Section~\ref{sec:discussion}, theoretical works indicate that the SFE increase may be less correlated with interactions prior to cosmic noon. Furthermore, as argued later, disk fragmentation in gas-rich environments at high redshifts could play a more significant role in increasing the galaxies' SFE.

Overall, we found that the fraction of interacting galaxies is increased in the starburst regime out to $z\sim4$ and correlates with SFE but not gas fraction (at least out to $z=2$). 
Galaxy-galaxy interactions therefore represent a veritable way to push galaxies into the starburst region at redshifts beyond cosmic noon.
However, our analysis also shows that interacting galaxies only make up at most $40\%$ of $z<3$ starburst galaxies {\it in our} sample. This implies that non-interacting systems contribute significantly to the starburst population.
The increase of star formation (efficiency) through disk instabilities could provide another avenue for galaxies to reach to starburst regime (see Section~\ref{sec:intro}). In this case, we would expect a higher fraction of disk galaxies with pronounced star-forming clumps in the starburst regime. This is studied in the next section.

\subsection{The Impact of Disk Instabilities on Starburst Activity}\label{sec:disk}

A significant fraction ($50-80\%$) of galaxies in the starburst regime in our sample are non-interacting ({\it i.e.} isolated) disk galaxies. These may have been interacting in the past, however, taking the current morphological evidence at face value shows that they are currently not in a merging state.
A possible way to reach the starburst regime is through an increased SFE due to the instability in gas-rich disks.
The idea \citep[see, for example,][]{dekel09,romeo16} here is that gas-rich streams increase the gas density of the disk, which then becomes unstable and starts to fragment into clumps \citep[Toomre instability;][]{toomre64}. These clumps can contribute to several percent of the total disk's mass and star formation is maintained in dense sub-clumps over timescales of several $100\,{\rm Myrs}$. Steady inbound gas streams on the disk maintain the instability of the disk and replenish gas over several Gyrs. Eventually, the clumps might migrate towards the center due to dynamical friction and may form spheroid-dominated galaxies later on.
Recent measurement with JWST using multiple broad and medium bands suggest that the majority of clumps are less than $\sim200\,{\rm Myrs}$ old, suggesting of a similarly long survival time \citep[e.g.,][]{claeyssens23}.

The occurrence of UV-bright star forming clumps in galaxies is ubiquitous at high redshifts \citep[about $60\%$ of $z=1-3$ galaxies contain bright UV clumps; ][]{cowie95,conselice04,elmegreen13,guo15,soto17,zanella19}.
Several studies support the formation of these clumps through {\it in situ} physical processes -- based on differences in the evolution of the clump fraction and minor/major mergers, the disk-nature of their host galaxies and similarly the kinematic properties of ordered disks rotation with high velocity dispersion, the distribution of clumps with scale height for edge-on galaxies, and their stellar mass function being similar to local star clusters and H~II regions \citep[see][]{shibuya16,elmegreen17,dessauges18,girard20}.
A recent study of lensed clumpy galaxies at $z=1$ \citep{dessauges23,dessauges19} corroborates the picture in which clumps are formed {\it in situ} through disk instabilities.
%

\begin{figure}[t]
\centering
\includegraphics[angle=0,width=1.03\columnwidth]{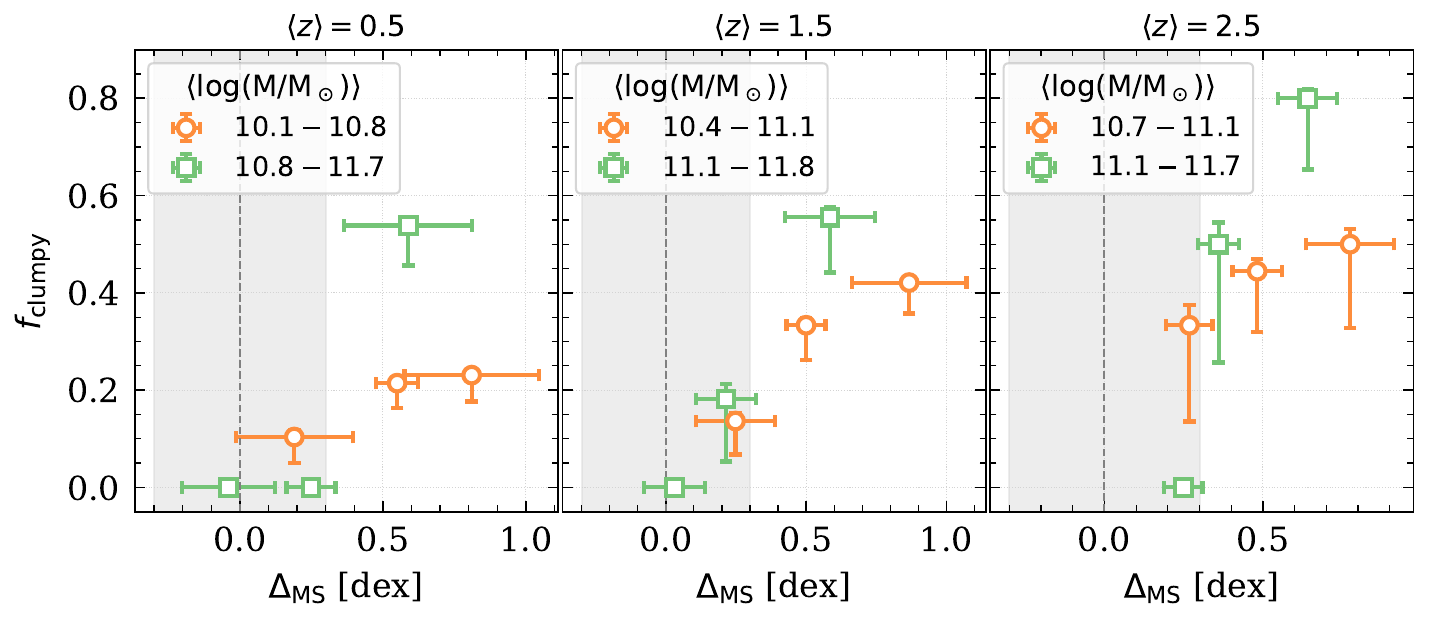}\\
\includegraphics[angle=0,width=1.03\columnwidth]{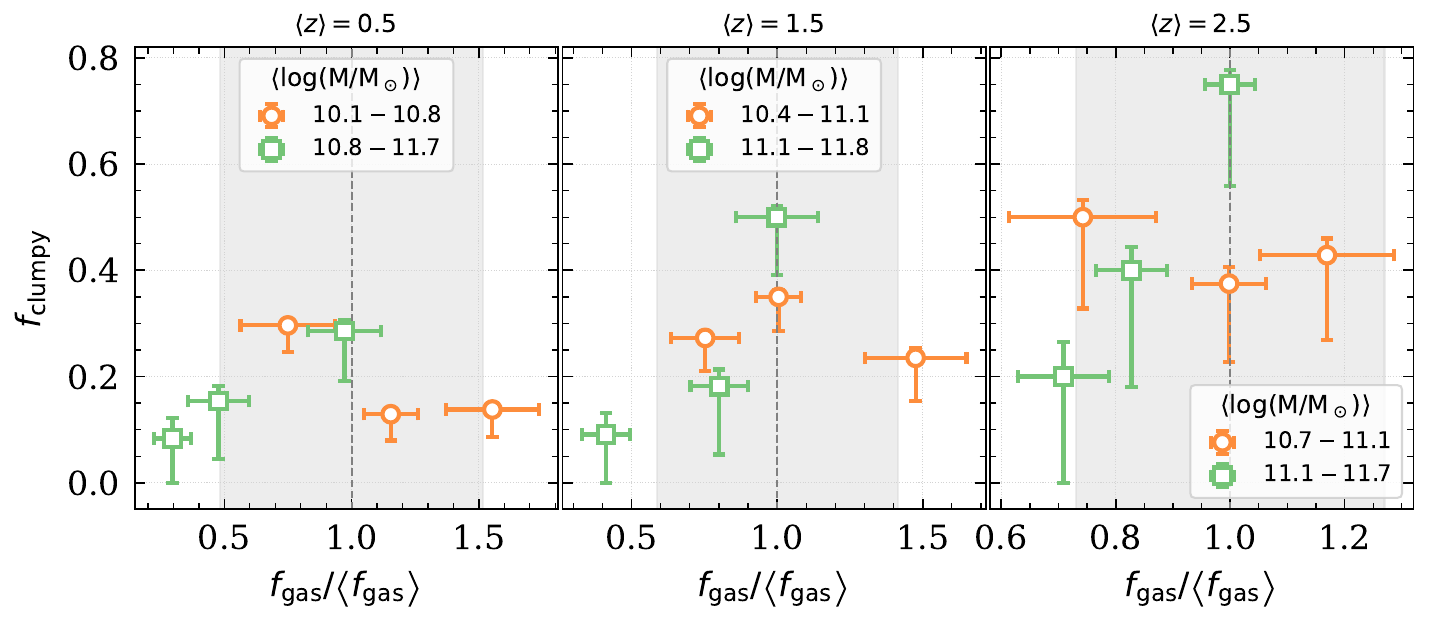}\\
\includegraphics[angle=0,width=1.03\columnwidth]{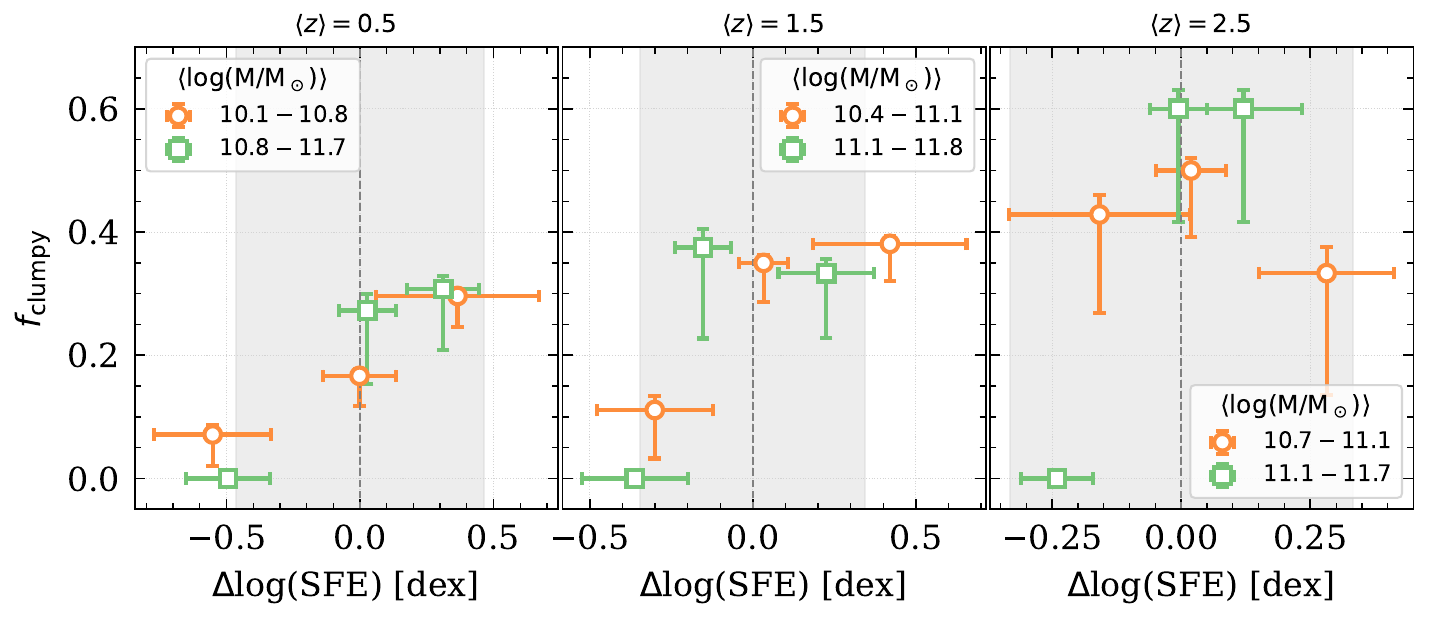}
\caption{Relation between the fraction of clumpy disks with respect to the total number of disks ($f_{\rm clumpy}$) and the offset from the main sequence ($\Delta_{\rm MS}$; top), gas fraction ($f_{\rm gas}$, middle), and star formation efficiency (SFE; bottom) for different redshift ranges ($\Delta z = 1$) and two stellar mass bins. The latter two quantities are normalized to the mean of the population (width indicated by the gray region) at a given redshift and stellar mass. The mass bins are chosen to mitigate selection biases. Note that we did not classify clumpy disk galaxies at $z>3$.
\label{fig:fclumpy}}
\end{figure}

Along these lines, \citet{wang94} presents a simple recipe to link star formation to disk instability. In this simple theoretical model, the SFE (as defined here as SFE/M$_{\rm gas}$) is proportional to:\vspace{-1mm}
\begin{equation}\label{eq:q}
    {\rm SFE} \equiv \frac{{\rm SFR}}{M_{\rm gas}} \propto \frac{(1-Q^2)^{1/2}}{Q}
\end{equation}
where $Q = \frac{\kappa v_{\rm gas}}{\pi G \Sigma_{\rm gas}}$ is the Toomre disk instability parameter, with $\kappa$ the epicyclic frequency, $v_g$ the radial cloud velocity dispersion, $\Sigma_{\rm gas} = \frac{M_{\rm gas}}{\pi r_{\rm disk}^2}$ the gas surface density, and $G$ the gravitational constant.
A similar expression can be derived from the analytical model presented in \citet[][see their equation 47]{dekel09}.
Equation~\ref{eq:q} is valid for an unstable disk, {\it i.e.} $Q<1$.
According to this model, the more unstable the disk, the more stars per surface are formed. Consequently, SFE rises as $1/Q$ as $Q\rightarrow 0$ (see Equation~\ref{eq:q}).
We emphasize the simplicity of this model \citep[see, for example, discussion in ][]{romeo11,romeo13,meidt23} as stars can also be formed if $Q>1$, however, $Q\ll1$ may be a condition of exceptional star formation.
Note, that by simply applying empirical correlations we get
\begin{equation}\label{eq:q2}
Q \propto \Sigma_{\rm gas}^{-1} \propto \frac{r_{\rm disk}^2}{M_{\rm gas}} \propto \frac{M_*^{0.4}}{\mu M_*} \propto M_*^{-0.24},    
\end{equation}
where we have used that $\mu = \frac{f_{\rm gas}}{1-f_{\rm gas}} \propto M_*^{-0.36}$ \citep{tacconi18} and $r_{\rm disk} \propto M_*^{\alpha}$ with $\alpha \sim 0.2$ \citep[e.g.,][]{yang21}. We therefore would expect only a weak stellar mass dependence.
In addition, we note that the dispersion $Q\propto v_g$ may be weakly positively correlated with stellar mass \citep[][]{ubler19}, making dependence of $Q$ on stellar mass even weaker.
If disk instabilities are at work to push disk galaxies into the starburst regime, we would expect an increase of galaxies with a number of dense star-forming clumps embedded in their disks for higher $\Delta_{\rm MS}$. That fraction would not significantly depend on stellar mass.

The top panel of Figure~\ref{fig:fclumpy} shows the fraction of clumpy disk galaxies (with respect to the isolated disk galaxy population) as a function of $\Delta_{\rm MS}$ for the same stellar mass bins as in Figure~\ref{fig:fint} and three redshift bins. We observe indeed an increase in the clumpy disk fraction towards the starburst regime.
In the center and lower panels of Figure~\ref{fig:fclumpy}, we provide a comparison of $f_{\rm clumpy}$ with gas fraction and SFE normalized to the population mean per bin. We find that the fraction of clumpy disks increases with both $f_{\rm gas}$ and SFE, however, the correlation with the latter is more pronounced. As expected, we do not find significant differences in that behavior for the different stellar mass bins (see Equation~\ref{eq:q2}), except for the comparison between $f_{\rm clumpy}$ and $f_{\rm gas}$ for which we do not find a significant correlation for the lower mass bin.
We find some weak trend of $f_{\rm clumpy}$ with redshifts (as shown by the different panels), which is in agreement with studies of larger galaxy samples \citep{guo15}.

\begin{figure}[t]
\centering
\includegraphics[angle=0,width=1.0\columnwidth]{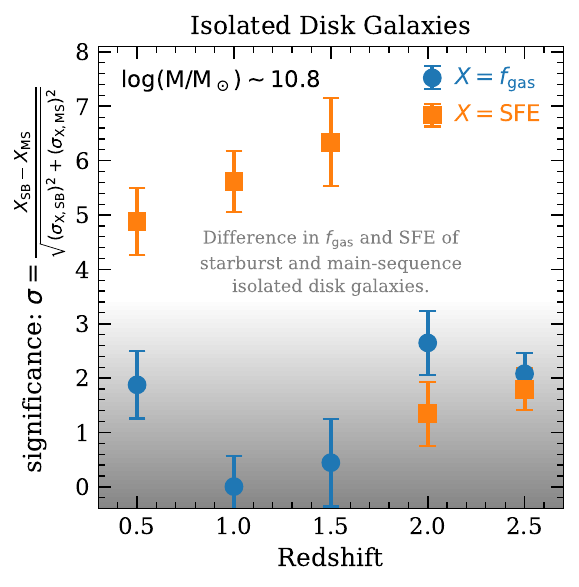}
\caption{Significance (in $\sigma$) of difference in $f_{\rm gas}$ (blue) and SFE (orange) between starburst and main-sequence disk galaxies out to $z=2.5$. For example, main-sequence and starburst disk galaxies at $z=0.5$ differ by $2\sigma$ and $5\sigma$ in $f_{\rm gas}$ and SFE, respectively, while at $z=2.5$, they differ at $2\sigma$ level. This shows that out to $z=2$, an increase in SFE is the dominant difference between disk galaxies on the main-sequence and in the starburst regime.
\label{fig:diskgalaxies}}
\end{figure}

Figure~\ref{fig:diskgalaxies} shows the above results more quantitatively by indicating how much $f_{\rm gas}$ and SFE differs (expressed in $\sigma$ values) between disk galaxies on the main sequence and in the starburst regime (defined as $>10\times$ above main sequence).
For this test, we mass-match the main-sequence and starburst disk sample to $\rm \log(M/M_\odot) = 10.8\pm0.3$ to remove potential dependence on stellar mass.
Averaged over redshifts up to $z=2$, the difference between $f_{\rm gas}$ of main-sequence and starburst galaxies is $<3\sigma$, while they differ by $>4\sigma$ in SFE.
This suggests that disk galaxies appear in the starburst population mainly due to an increased SFE (achieved through disk instabilities) and not an increase in the gas fraction. An increase in $f_{\rm gas}$ may contribute at higher redshifts as the difference in SFE and $f_{\rm gas}$ becomes comparable between two types of galaxy populations.

\section{Discussion \& Conclusions} \label{sec:discussion}

In this work, we have collected a sample of galaxies to investigate the properties of starbursts out to $z~\sim~4$ in terms of their morphology, gas fraction, and star formation efficiencies.
Our sample is well fit for this investigation, as all galaxies have robust total (UV$+$far-IR) star formation rates, gas mass measurements, and high-resolution optical to near-IR imaging from JWST.
In summary, these are our two main findings:
\begin{itemize}
    \item We find a significant enhancement of the fraction of interacting galaxies in the starburst regime out to $z\sim4$. This fraction correlates with SFE (but less with $f_{\rm gas}$) out to $z\sim2$, but no significant correlation is found between $2<z<4$.
    \item A large fraction of starburst galaxies in our sample are non-interacting disks ($50-80\%$ depending on redshift). We find that the fraction of clumpy disk galaxies with respect to the disk population is enhanced in the starburst regime. This increase correlates more significantly with SFE than $f_{\rm gas}$.
\end{itemize}

We note that with our visual classification we are most sensitive to major to equal-mass mergers. This is suggested by the flux ratios of the merging components in the red filters, which are tracing their stellar mass. As mentioned earlier, kinematic spectroscopic measurements and deeper imaging data would likely be necessary to identify minor mergers robustly. The here identified interacting systems therefore do not contribute to the more continuous accretion of mass and gas as minor mergers would do \citep[e.g.,][]{scoville23}.

The former of these findings suggests that galaxy-galaxy interactions are a way to push galaxies into the starburst regime by increasing the efficiency of star formation in the galaxies.
This has been observed at lower redshifts and is suggested by simulations \citep{sanders96,genzel10,kartaltepe12,hung13,hopkins18,cibinel19,renaud19,moreno21,wilkinson22,he23,shah22}.
Specifically, we showed that the fraction of interacting galaxies increases with increasing $\Delta_{\rm MS}$ out to $z\sim4$. However, we also find that the interacting fraction does correlate with SFE only out to $z=2$, while the correlation becomes weaker in the earlier universe and lower stellar masses. This is also indicated by some simulation results \citep[e.g.,][]{fensch17,patton20}, and could suggest a weaker correlation between interaction and SFE increase, or that such a correlation occurs on a shorter timescale, before the cosmic noon. It also suggests that galaxy-galaxy interactions may not be the prime avenue to cause starbursts at $z>2$ -- instead (as shown in Section~\ref{sec:disk}), disk instabilities could be a more common path as $f_{\rm clumpy}$ clearly correlates with $\Delta_{\rm MS}$ out to at least $z=4$.

We note that even in the case of a strong 1-to-1 correlation between $f_{\rm int}$ and SFE, this signal could be diluted leading to no significant correlation between these two quantities.
The main reason for this is the different time scales probed here. Specifically, we would expect an increase in star formation (and hence SFE assuming little changes in the gas mass) triggered through the compression of gas upon close interaction within a few $100\,{\rm Myrs}$ \citep{lotz08,teyssier10,cenci23}. Star formation traced by optical emission line indicators (e.g., \halpha) may provide the closest link between SFE increase and close interactions. In fact, simulations show that {\it infrared} SFR tracers can lag behind as far-IR light is emitted mostly after coalescence when sufficient amounts of metals and dust have been produced \citep[e.g.,][]{lotz08}. Therefore, while a strong increase in SFE can be observed in close mergers, the late and early-stage mergers (averaging over many $100\,{\rm Myrs}$) selected in our classification may not directly trace a SFE increase \citep[see also][]{teyssier10,shah22}. Thus averaged over the whole sample of interacting galaxies, the correlation between $f_{\rm int}$ and SFE can be significantly diluted due to the mismatch in time scales and the short time-span of increased SFE (or likewise SFR) during interactions.

The significant fraction of non-interacting disk galaxies in our sample defined as starbursts indicates another avenue other than galaxy-galaxy interactions at work to offset galaxies from the main sequence \citep[cf. starburst rotators found at $z=2$, e.g.,][]{forsterschreiber11}. Other studies \citep[e.g.,][]{wilkinson22} have found similar fractions, and in this work we have studied the properties of these galaxies.
As shown by Equation~\ref{eq:q} and also by recent simulation work \citep{cenci23}, disk instabilities can cause an increase in SFE, which can push galaxies above the main sequence. We expect that this effect may be largely independent of stellar mass (Equation~\ref{eq:q2}).
The enhanced fraction of clumpy disk galaxies in the starburst population of isolated disk galaxies in our sample and the coupling between $f_{\rm clumpy}$ and SFE are direct observational indications of disk instabilities in action.

It is interesting that while the fraction of clumpy disks shows a significant correlation with SFE, the fraction of interacting galaxies does not. This could suggest that the SFE increase due to disk instabilities is maintained over longer timescales (compared to merger-driven enhancement as discussed above), thus it is statistically more easy to observe a correlation. The factor of $\sim3$ higher fraction of galaxies with clumps at high $\Delta_{\rm MS}$ compared to interacting galaxies is an additional indication of a longer sustained increase of SFE through disk instabilities.
As mentioned above, recent JWST observations suggest that the clump survival time is likely less than $200\,{\rm Myrs}$. If the SFE increase is indeed maintained over longer timescales, this would mean that the gas in the disk needs to be replenished on similar time-scales as the clump lifetime in order to maintain disk fragmentation.

More detailed analyses, including kinematic measurements with JWST or ALMA, are necessary to study the effect of interactions on starburst galaxies in more detail. Specifically, \halpha~observations would provide an important constraint on the more instantaneous SFE. At higher redshifts, a statistical study of star forming clumps in starburst disk galaxies would reveal their properties, specifically their ages and star formation densities, similarly to what has been achieved at $z=2$ by multiple IFU programs \citep[e.g.,][]{forsterschreiber20}.

Further characterization of star-forming clumps in larger galaxy samples will additionally become important in understanding the role of disk instabilities in growing the most massive galaxies at the highest redshifts. A possible model to explain the high star formation efficiencies needed to produce the most massive galaxies during the Epoch of Reionization \citep[e.g.,][]{harikane23,xiao23,casey23b} are localized regions of essentially feedback-free star formation \citep[e.g.,][]{dekel23,lizhaozhou23}. This model specifically works in low-metallicity regions that are self-shielded from significant external stellar winds. In such a regime, stars are efficiently formed before the feedback introduced by supernovae starts \citep[for more details see][]{dekel23}.
A way to form these dense pockets of star formation could be through violent disk instabilities in gas-rich galaxies fed by pristine low-metallicity gas. 
Interestingly, such a scenario may have been recently observed in a lensed $z=6$ galaxy hosting $15$ star-forming clumps \citep{fujimoto24}. The high gas surface density of the clumps ($\Sigma_{\rm gas} \sim 10^{3-5}\,{\rm M_\odot\,pc^{-2}}$) suggests an overall star formation efficiency (assuming star formation takes place in the clumps) that is $6-9\times$ higher \citep{fukushima21} than for typical main-sequence galaxies with densities $\Sigma_{\rm gas} < 10^{3}\,{\rm M_\odot\,pc^{-2}}$, assuming star formation takes place in their extended disks \citep[e.g.,][]{dessauges20,tacconi18}.
The contribution of the clumps to the integrated star formation is expected to be significant ($>50\%$) in this case, in contrast of other studies finding contributions in the low tens of per-cent in galaxies after the cosmic noon \citep{elbaz18,cibinel17,hodge16}.
The integrated $f_{\rm gas}$ of that galaxy is $\sim 0.75$, thus not significantly higher than main-sequence galaxies at a similar redshift \citep[$0.4-0.8$,][]{dessauges20}.
If disk instabilities play an important role in increasing the star formation efficiency (as shown by our work and recently by \citealt{fujimoto24}), the most massive galaxies in the early universe (although they look compact) may show similar structures on resolved to scales of $\sim100\,{\rm pc}$ \citep[unless other formation channels are present such as monolithic collapse, e.g.,][]{eggen62}. Such scales can currently only be probed by gravitational lensing, however, later this may be possible with diffraction limited observations using the next generation of large telescopes such as the {\it Thirty Meter Telescope} (TMT) or the European {\it Extremely Large Telescope} (ELT).

\begin{acknowledgments}
The JWST data presented in this article were obtained from the Mikulski Archive for Space Telescopes (MAST) at the Space Telescope Science Institute. The specific observations analyzed can be accessed via\dataset[DOI]{https://doi.org/10.17909/ph8h-qf05}.

We thank the anonymous referee for their suggestions, which substantially improved this manuscript.
Support for this work was provided by
NASA grants JWST-GO-01727 and HST-AR15802 awarded by the Space Telescope Science Institute, operated by the Association of Universities for Research in Astronomy, Inc., under NASA contract NAS 5-26555.
CMC thanks the National Science Foundation for support through grants AST-1814034 and AST-2009577 as well as the University of Texas at Austin College of Natural Sciences for support.
CMC also acknowledges support from the Research Corporation for Science Advancement from a 2019 Cottrell Scholar Award sponsored by IF/THEN, an initiative of Lyda Hill Philanthropies.
The French part of the COSMOS team is partly supported by the Centre National d'Etudes Spatiales (CNES).
OI acknowledges the funding of the French Agence Nationale de la Recherche for the project iMAGE (grant ANR-22-CE31-0007).
This work was made possible by utilizing the CANDIDE cluster at the Institut d'Astrophysique de Paris. The cluster was funded through grants from the PNCG, CNES, DIM-ACAV, the Euclid Consortium, and the Danish National Research Foundation Cosmic Dawn Center (DNRF140). It is maintained by Stephane Rouberol.
SG acknowledges financial support from the Villum Young Investigator grant 37440 and 13160 and the Cosmic Dawn Center (DAWN), funded by the Danish National Research Foundation (DNRF) under grant DNRF140.
\end{acknowledgments}

%

\facilities{ALMA, Herschel, JWST}


\software{
\texttt{astropy} \citep{astropy13,astropy18};
\texttt{PhotUtils} \citep{photutils_bradley_2024};
\texttt{SExtractor} \citep{bertin96};
\texttt{statmorph} \citep{rodriguezgomez19};
\texttt{WebbPSF} \citep{perrin14}
}



\appendix
\counterwithin{figure}{section}

\section{Statistical Morphological Classification} \label{sec:statmorph}

For this work, we performed a visual classification of our galaxies in mainly disks, clumpy disks, and interactive systems. Here, we compare this visual classification with a statistical morphological classification. Specifically, the Gini ($G$) vs. $M_{\rm 20}$ \citep[e.g.,][]{lotz04} or concentration ($C$) vs. asymmetry ($A$) diagnostics \citep[e.g.,][]{bershady00} diagnostics are some of the most used classifiers of interacting and non-interacting galaxy systems.

We use \texttt{statmorph} \citep{rodriguezgomez19} to measure these non-parametric morphological quantities. We first create a mask for each source to remove contaminants. A mask for each filter was generated using the {\it detect\_sources} function provided by the \texttt{PhotUtils} \citep{photutils_bradley_2024} Python package and requiring a minimum of 20 connected pixels and a threshold of $>2.5\,\sigma$ above background. The masks are then carefully combined to a ``master mask'' to optimally enclose the flux of the source and to remove contaminants. We then run \texttt{statmorph} on all galaxies for all their available bands using the masks as created above.

\begin{figure}[t]
\centering
\includegraphics[angle=0,width=1.0\columnwidth]{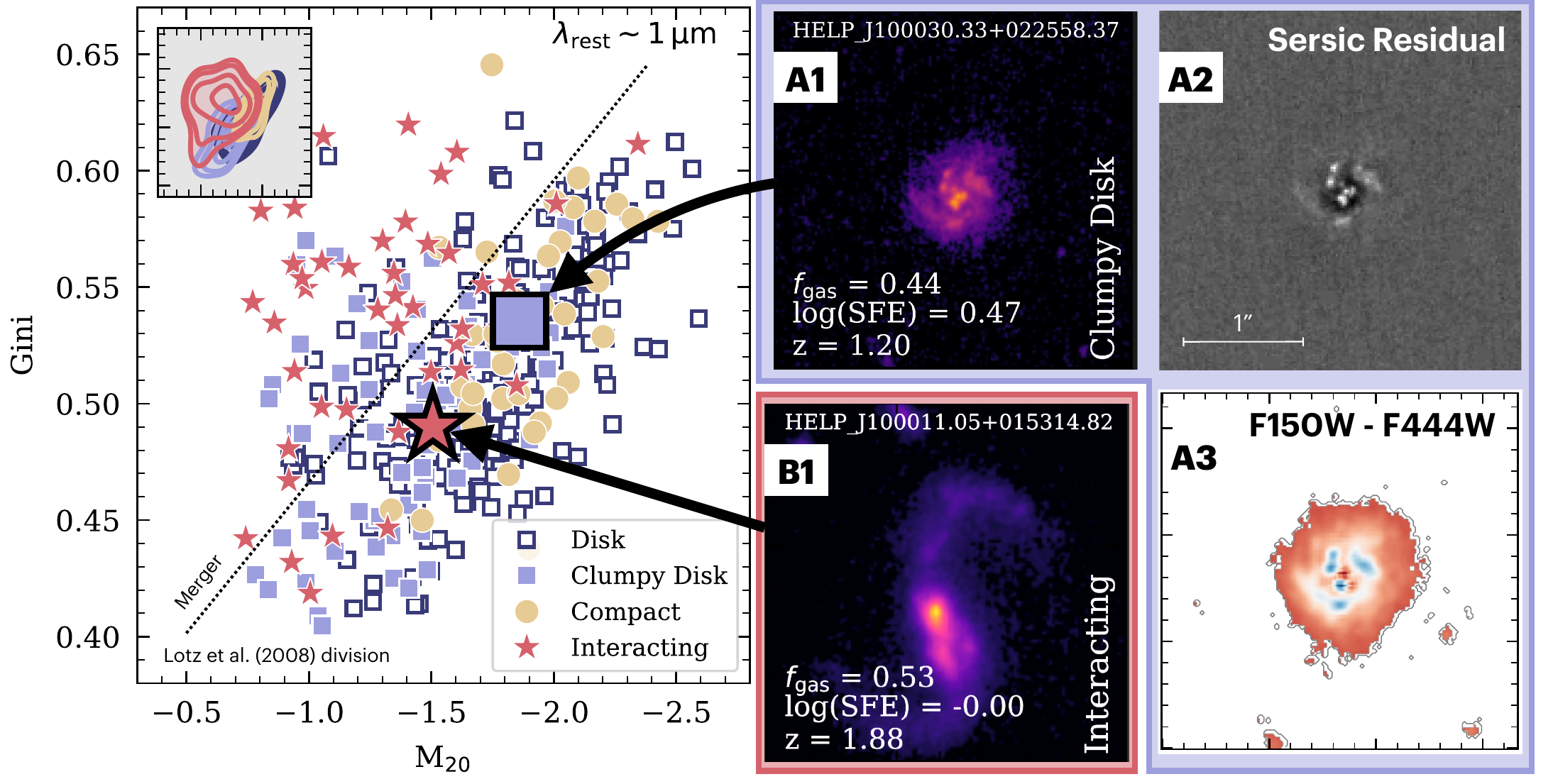}
\caption{On the statistical morphological classification compared to visual classification.
{\it Left Panel}: Gini vs. M$_{\rm 20}$ diagram with visually classified {\it Disks} (dark blue open squares), {\it Clumpy Disks} (light blue squares), {\it Compact} (yellow circles), and {\it Interacting} (red stars) systems (see Section~\ref{sec:visclass}). The division lines between interacting and disk systems from \citet{lotz08} is indicated as dashed line. The completeness of the selection of interacting galaxies using this method is high, however, the purity is low (specifically many of the interacting systems are classified as clumpy disks). The small inset shows Gaussian kernel estimations for the different visual classes.
{\it Right Panels}: Example of a visually classified interacting systems (indicated by the tidal tail structure ({\it B1}), that would be missed by an statistical selection. Panels {\it A1-3} show a visually classified clumpy disk (F150W, Sersic residual, and color map image). Statistical methods such as $G-M_{\rm 20}$ are not suitable to identify clumpy disks for this work.
\label{fig:autoclass}}
\end{figure}

The left panel of Figure~\ref{fig:autoclass} shows the resulting $G-M_{\rm 20}$ diagram for the galaxies at rest-frame $\sim1\,{\rm \mu m}$. A commonly used differentiation between merging and non-merging galaxies \citep[e.g.,][]{lotz08} is indicated as a dashed line.
It can be seen that $60-70\%$ of visually classified interacting systems are recovered. This number can be improved slightly by optimizing the statistical selection criteria. However, while the completeness is high, the purity is not. Specifically, as shown in the figure, the clumpy disk galaxies occupy a similar space as the interacting systems. 
On the other hand, some interacting systems would not be classified as mergers (e.g., Panel $B1$ in Figure~\ref{fig:autoclass}). We notice that these are usually interacting galaxies identified visually by their tidal tail structures. These may be mergers in final approach.
We also investigated the $C-A$ classification diagram \citep{bershady00}, however, we found that the separation between interacting and non-interacting is less clear and therefore we do not go into further details here.

The panels $A1-3$ of Figure~\ref{fig:autoclass} show an example of a clumpy disk galaxy. It would not be identified as such using the $G-M_{\rm 20}$ statistical classification as it is too similar to an isolated disk galaxy. Panel {\it A1} shows the F150W image. Panel {\it A2} shows the same image but with subtracted Sersic model fit by \texttt{statmorph}, revealing the disk structure and clumps in the residual. Panel {\it A3} shows the F444W image subtracted from the PSF-matched F150W image. The PSF matching was achieved with the \texttt{PhotUtils} function \textit{create\_matching\_kernel()} using the \texttt{WebbPSF} \citep{perrin14} theoretical PSFs of the two bands. The star forming clumps are clearly identifiable in the Sersic-subtracted residual image and the color image (as blue clumps). This is an example of the technique used for visual classification and identification of clumpy disk galaxies. A more in-depth study of these clumps and their identification will be provided in a follow-up paper (Kalita et al. in prep.).

All in all, this analysis shows that our visual classification of interacting systems is robust. Furthermore, we believe that a statistical morphological identification would not be as reliable as our visual classification by members of our team, given the difficulties in selecting clumpy disk galaxies. This is especially true for such a relatively small sample of galaxies. A more sophisticated treatment of this selection using machine learning algorithms is therefore also beyond the scope of this work.


\bibliography{bibli}{}
\bibliographystyle{aasjournal}



\end{document}